\documentclass[aps,prd,reprint,preprintnumbers,nofootinbib]{revtex4-2}
\usepackage{physics}
\usepackage{graphicx} 
\usepackage{amsmath,amssymb,amsthm,amsfonts}
\usepackage[normalem]{ulem}
\usepackage{comment}
\usepackage{dsfont}
\usepackage[svgnames]{xcolor}
\usepackage[
  colorlinks=true,
  linkcolor=DarkRed,
  citecolor=DarkGreen,
  urlcolor=DarkBlue
]{hyperref}

\usepackage{natbib}

\newcommand{\Sp}[1]{\text{span}{(#1)}}

\begin{document}
\preprint{LA-UR-25-32226}
\title{Efficient Learning of Lattice Gauge Theories with Fermions}

\author{Shreya Shukla}
\thanks{First author; remaining authors are listed in reverse alphabetical order}
\email{\\sshukla@lanl.gov}
\author{Yukari Yamauchi}
\email{yyamauchi@lanl.gov}
\author{Andrey Y. Lokhov}
\email{lokhov@lanl.gov}
\author{Scott Lawrence}
\email{srlawrence@lanl.gov}
\author{Abhijith Jayakumar}
\email{abhijithj@lanl.gov}
\affiliation{Theoretical Division, Los Alamos National Laboratory, Los Alamos, NM 87545}

\begin{abstract}
    We introduce a learning method for recovering action parameters in lattice field theories.
    Our method is based on the minimization of a convex loss function constructed using the Schwinger-Dyson relations.
    We show that score matching, a popular learning method, is a special case of our construction of an infinite family of valid loss functions. Importantly, our general Schwinger-Dyson-based construction applies to gauge theories and models with Grassmann-valued fields used to represent dynamical fermions. In particular, we extend our method to realistic lattice field theories including quantum chromodynamics.
\end{abstract}
\maketitle

\section{Introduction}
Lattice field theory provides systematic, non-perturbative access to a wide range of quantum systems, most famously including quantum chromodynamics (QCD), the theory of strong interactions. Given a family of lattice field theories (such as lattice QCD), there are two computational tasks one might perform. Most commonly, we fix a set of parameters and couplings, and compute expectation values over the probability distribution thus defined. The inverse task can also be defined: given a set of expectation values, what lattice action was used to compute them?

This inverse task, sometimes called the learning problem, is the subject of this work. The learning problem is not a mere mathematical curiosity: it is motivated by a number of applications. One of the central objects in quantum field theory is the renormalization flow. By sampling at one set of parameters, coarse-graining, and then solving the learning problem, we can directly study how the coupling constants change under renormalization~\cite{Shanahan:2018vcv,Shukla:2025wze}.

In principle, one can solve this learning problem by repeatedly performing the forward procedure, for different coupling constants, until the expectation values are found to match. This is inefficient for two reasons. First, computing expectation values is computationally difficult; second, the space of possible couplings can be high-dimensional and therefore expensive to search through. This paper is motivated by the desire for a more efficient approach. Indeed, our approach is inspired by an observation that the learning problem can often be solved \emph{more} efficiently than the forward problem~\cite{lokhov2018optimal}.

In the case of certain statistical models, the learning problem is efficiently solved by a popular method known as \emph{score matching}~\cite{Hyvarinen2005} that avoids the computation of the partition function. Standard score matching may be applied to lattice field theories with continuous variables, as was done in~\cite{Shukla:2025wze}, but encounters computational difficulties in the case of Grassmann variables that are used to model fermions in important physical theories. Moreover, score matching requires observation of a specific set of correlators that may not be conveniently obtained, and might not be the most data-efficient estimator. This motivates a search for a more general family of estimators that would would conveniently apply to relevant lattice gauge theories.

Our work begins with the observation that a set of equations known as Schwinger-Dyson relations can be used as a foundation for a learning method. The Schwinger-Dyson relations are a set of constraints on the expectation values of a statistical system, which can be efficiently determined by inspection of the action, with only a trivial amount of computation. These relations have been used before as a method of variance reduction~\cite{Lawrence:2024xsi,Bhattacharya:2023pxx,Bedaque:2023ovz}; in this work we show how they may be used as the basis of a learning algorithm. In Section~\ref{sec:two}, we show how to construct a broad family of efficient loss functions based on suitable subset of Schwinger-Dyson relations. Interestingly, we show that score matching is a special case of our construction corresponding informed by the sufficient statistics of the model. Although we leave detailed analysis of sample and computational complexity to future work, we demonstrate in Appendix~\ref{app:not-optimal} that score matching is in general not the most efficient member of this family of estimators.

While score matching is not obviously applicable to theories of Grassmann variables, Schwinger-Dyson equations do exist for such theories. In Section~\ref{sec:grassmann} we show that these Schwinger-Dyson relations may be used for learning of action parameters given expectation values. We demonstrate this approach on a lattice Thirring model.

Arguably the most important lattice field theories, including lattice QCD, are gauge theories. In Section~\ref{sec:gauge} we discuss the interaction of Schwinger-Dyson-based learning and score matching with gauge symmetry. We further detail how these learning methods may be used in the case of full-fledged lattice QCD. Finally in Section~\ref{sec:discussion} we consider the prospects for, and limitations of, these learning methods.

\section{Learning method}\label{sec:two}
In this work we are concerned with a particular type of family of probability distributions: exponential families. These are families of probability distributions $p(x)$ that have the form
\begin{equation}
    p(x) = \frac{1}{Z} e^{-S_0(x) - \sum_i \theta^{*}_i Q_i(x)}\text,
    \label{eq:generic_exp_family}
\end{equation}
for known functions $S_0$ and $Q_i$ and potentially high-dimensional $x$, and $Z$ is the normalization often referred to as the partition function. Such families are widespread in both statistical and quantum field theories. The coefficients $\{\theta^{*}_i\}$ (collectively denoted as $\theta^{*}$) are termed the couplings and specify a member of the exponential family. In the context of field theory there is an additional condition: the functions $S_0$ and $Q_i$ are polynomials in the ``fields'' $x$.

The Schwinger-Dyson relations~\cite{Dyson1949,Schwinger1951a,Schwinger1951b} are an infinite tower of linear constraints on expectation values, commonly used in the context of quantum field theory. In this section, we show how they can be used as a basis for a learning method. We also show that score matching~\cite{Hyvarinen2005}, a widely used method in machine learning literature to solve the learning problem associated with the exponential family of distributions, emerges as just one choice among the infinite family of loss functions which can be constructed by considering the Schwinger-Dyson relations.

As simple illustration of the main ideas in this work, consider the following system, which may be considered a $0$-dimensional $\phi^4$ theory with expectation values defined as follows:
\begin{equation}\label{eq:ii-example}
    \langle f(x) \rangle = \int dx\, f(x) e^{-\frac{\alpha^{*}}{2} x^2 - \frac{\lambda^{*}}{4} x^4}
    \text,
\end{equation}
for scalar $x$. The learning problem is as follows: given a such few expectation values (perhaps $\langle x^2\rangle$ and $\langle x^4\rangle$), determine the values of the ``couplings'' $\alpha^{*}$ and $\lambda^{*}$. For example, if we are promised $\lambda^{*} = 0$, this is particularly easy, as we may immediately calculate $\langle x^2\rangle = (\alpha^{*})^{-1}$.

In the case where $\lambda^{*}$ does not vanish, expectation values of the form $\langle x^{2k}\rangle$ are not straightforwardly expressed as a function of the couplings. However, a subset of expectation values are analytically related to each other by the couplings, and we can use these to determine the unknown couplings that define the theory, as we show below.

\subsection{Learning from Schwinger-Dyson relations}

Analytical relations between expectation values are termed Schwinger-Dyson relations, and are easily obtained via integration by parts. For any distribution over $\mathbb R$ of the form $e^{-S(x)}$ (with polynomial $S(x)$ of even degree $\ge 2$), and any polynomial $P(x)$, observe that
\begin{equation}
    \int_{-\infty}^{+\infty} dx\, \frac{d}{d x} \left(P(x) e^{-S(x)}\right) = 0
    \text.
\end{equation}
This expression generalizes to higher-dimensional spaces (like $\mathbb R^N$) and compact integration domains (like $SU(N)$). It immediately follows that, for any first-derivative operator $\partial$ and any function $P(x)$ such that $e^{-S(x)} P(x)$ is sufficiently quickly decaying, we have the following linear constraint on the expectation values:
\begin{equation}\label{eq:sd-1d}
    \langle P(x) \partial S(x)\rangle = \langle \partial P(x) \rangle
    \text.
\end{equation}

In the case of our $0$-dimensional $\phi^4$-like example, the general Schwinger-Dyson relation reads
\begin{equation}
    k \langle x^{k-1}\rangle = \alpha^{*} \langle x^{k+1}\rangle + \lambda^{*} \langle x^{k+3}\rangle\text.
\end{equation}
When the parameters $\alpha^{*}, \lambda^{*}$ are known, this provides a useful constraint on the expectation values. In our case, the expectation values are known but the action parameters are not. The above equation makes clear that knowing $\langle x^2 \rangle$, $\langle x^4\rangle$, and $\langle x^6\rangle$ is sufficient to determine $\alpha^{*}, \lambda^{*}$ by linear algebra.

The remainder of this paper will consist of variations and elaborations on this theme. We are provided (perhaps stochastic estimates of) expectation values of some theory. We know the action of the theory has a particular form, but the coupling constants in this action are not known. We wish to determine the couplings. In order to do this, we write down a sufficiently large set of Schwinger-Dyson relations. We find the couplings that would minimize the amount of violation of these Schwinger-Dyson relations, and these couplings are our estimate of the couplings that were used to produce the original observables.

Let us formalize this procedure for the generic model of the type \eqref{eq:generic_exp_family}. First, for a choice of $n$ test polynomials $P_1, \ldots, P_n$, define the residual vector
\begin{align}
    r_j(\theta) &= \langle \partial P_j\rangle - \langle P_j \partial S\rangle\;,
\end{align}
where $S = S_0 + \sum_k \theta_k Q_k$ denotes action parametrized by the optimization parameters $\theta$. This residual vanishes when evaluated at the true parameters: $r_j(\theta^*) = 0$. It will also be convenient to define
\begin{equation}
    M_{ji} = \langle P_j \partial Q_i \rangle
    \text{ and }
    b_j = \langle \partial P_j\rangle -\langle P_j \partial S_0\rangle
    \text.
\end{equation}
In this notation the residual is defined by $r_j(\theta) = b_j - \sum_i M_{ji}\theta_i$.

With the residual defined, we can construct a convex loss function from any positive-definite quadratic form. Concretely, for any positive definite symmetric matrix $A$ we define
\begin{align}\label{eq:LA}
    \mathcal{L}_A(\theta) = \sum_{i,j}r^T_i(\theta)A_{ij} r_j(\theta)
    \text.
\end{align}
Minimization of the loss function defined in \eqref{eq:LA} for a certain choice of test polynomials $\{P_i\}$ and positive definite matrix $A$ constitutes our proposed learning method for lattice gauge theories. When provided with exact expectation values---that is, when there are no statistical errors---this loss function is minimized precisely when $r=0$, at which point the true couplings $\theta^{*}$ are recovered. In the case of finite samples, the Schwinger-Dyson equations will not be satisfied exactly, but the $\widehat{\theta}$ which minimizes $\mathcal L_A$ results in an estimator for the true parameters $\theta^{*}$. The estimator properties will crucially depend on the choice of both $\{P_i\}$ and $A$.

Suppose that we have chosen some basis of polynomials $P_1,\ldots,P_n$. Under what circumstances is the knowledge of the corresponding Schwinger-Dyson expectation values sufficient to uniquely fix the parameters $\theta^{*}$? As we discuss below, the choice $P_i = \partial Q_i$ is sufficient for estimating the model parameters (and any choice which is a superset will also suffice), but how much freedom in the choice of test polynomials $\{P_i\}$ is available? Intuitively, as long as the $\{P_i\}$ are independent and ``generic'' with respect to the known lattice symmetries\footnote{In the one-dimensional example above, note that the Schwinger-Dyson relations generated by $P_k = x^{2k}$ provide no information whatsoever about $\lambda^*$, precisely because of the $\mathbb Z_2$ symmetry.}, and there are at least as many of them as there are free parameters, the parameter estimates $\widehat{\theta}$ should be uniquely determined.

We can understand geometrically the conditions under which the parameters $\theta^*$ are uniquely fixed, as follows. We begin by defining an inner product on the space of polynomials:
\begin{equation}
    \langle f,g \rangle_p = \int dx\,p(x) f(x) g(x)\text.
\end{equation}
This inner product also defines the orthogonal space $\mathcal V^\perp$ of a vector space $\mathcal V$. With these definitions we observe that the action of the matrix $M$ may be written in terms of this inner product:
\begin{equation}\label{eq:M-action}
    (M v)_i = \big\langle P_i, \sum_j \partial Q_j v_j\big\rangle_p
\end{equation}
Assuming positive-definite $A$, the minimum of the loss function Eq.~(\ref{eq:LA}) is unique if and only if $\ker M = \{0\}$. By inspection of Eq.~(\ref{eq:M-action}), the kernel of $M$ is precisely those vectors $v$ which produce linear combinations $v_j \partial Q_j$ which are orthogonal to all linear combinations of the $\{P_i\}$. The kernel vanishes if and only if there is no linear combination of the $\{\partial Q_j\}$ which yields a function orthogonal to all $P_i$. Thus a condition both necessary and sufficient for $\theta^*$ to be uniquely determined is:
\begin{equation}\label{eq:necessary-sufficient}
    \Sp{\{P_i\}}^{\perp} \cap \Sp{\{\partial Q_j\}} = \{0\}\text.
\end{equation}
Assume that this condition does not hold, i.e., there exists $q \in \Sp{\{P_i\}}^{\perp} \cap\, \Sp{\{\partial Q_j\}}$ such that $q \neq 0$. Since $q \in \Sp{\{\partial Q_j\}}$, there exists $v^* \neq 0$ such that $q = \sum_j v^*_j \partial Q_j$, implying that $\langle P_i,\sum_j v^*_j\partial Q_j\rangle_p = 0$ for all $i$ and $v^* \in \ker{M}$. For a solution $\theta_0$ of the Schwinger-Dyson equations, we have for any $t \in \mathbb{R}$
\begin{align}
    M(\theta_0 + tv^*) &= b + 0 = b\;,
\end{align}
and $v^*$ manifests as flat direction in the loss function Eq.~(\ref{eq:LA}), implying that we no longer have a unique minimum.


Note that the condition Eq.~(\ref{eq:necessary-sufficient}) cannot in general be checked until expectation values are known, as the definition of $\mathcal V^\perp$ depends on the expectation values with respect to $p$.

\subsection{Relation to score matching}

Minimization of the loss function defined in \eqref{eq:LA} for any sufficient set of test polynomials $\{P_i\}$ discussed above and any positive definite matrix $A$ yields parameter estimates $\widehat{\theta}$ that converge to $\theta^{*}$ in expectation. Here, we discuss an important special case for the choice of $\{P_i\}$ and $A$ that recovers a widespread learning method for estimating probability distributions with continuous variables: score matching~\cite{Hyvarinen2005}. The loss function of score-matching corresponds to a choice of test polynomials which is guaranteed to be sufficient, and in part for this reason, we will use score matching as the basis of our investigation of gauge theories in Section~\ref{sec:gauge}. 

We start by providing a quick background on score matching. As before, assume that expectation values with respect to an exponential family distribution $p(x)$ of the type \eqref{eq:generic_exp_family} are known, but parameters $\theta^*$ are unknown. Define an ansatz distribution $\pi(x;\theta)$ in the same family, parameterized by a vector $\theta$.
To find $\theta^*$, we minimize the following \emph{score-matching} loss function over the parameters $\theta$:
\begin{equation}
    \mathcal L(\theta) = \Big\langle
    \left\Vert
    \frac{\partial}{\partial x} \log p(x) 
    -
    \frac{\partial}{\partial x} \log \pi(x;\theta)
    \right\Vert^2
    \Big\rangle_{p(x)}
    \text,
    \label{eq:score_matching_loss}
\end{equation}
where $\frac{\partial}{\partial x} \log p(x)$ is often termed \emph{score function} of the distribution $p(x)$. In the score matching loss function, the expectation value is taken over the empirical probability distribution $p(x)$. In the multidimensional case, $\frac{\partial}{\partial x}$ denotes the vector-valued gradient, and $||\cdot||$ denotes the $L_2$-norm of the vector of partial derivatives. 
Notice, using the fact that $p$ vanishes nowhere, that $\theta^*$ is the unique minimum (and in fact the unique zero) of the loss function $\mathcal L(\theta)$.

As written, it is not practical to minimize this loss, because evaluating $\mathcal L(x)$ requires direct knowledge of $p(x)$. The fact that expectation values are taken over $p(x)$ is not an obstacle---we know such expectation values by assumption---but the fact that the precise expectation value to be taken explicitly depends on $\log p(x)$, is. However, with some algebraic manipulation (see \cite{Hyvarinen2005} for details), we can rewrite the same loss function, up to a $\theta$-independent constant, as
\begin{equation}\label{eq:loss_estim}
\mathcal L(\theta) = \Big\langle \Big(\frac{\partial}{\partial x}S_\theta\Big)^2\Big\rangle - 2 \Big\langle \frac{\partial^2}{\partial x^2} S_\theta \Big\rangle
    \text.
\end{equation}
Here we have labeled by $S_\theta$ the logarithm of the ansatz distribution $\pi(x;\theta)$. This defines the score matching loss used in practice: this form of the loss function is readily evaluated, given samples over the empirical distribution $p(x)$.

Let us illustrate the loss \eqref{eq:loss_estim} with our toy example of the $0$-dimensional $\phi^4$ theory, with the trial distribution $\pi(x;\alpha,\lambda) \propto e^{-\frac{\alpha} 2 x^2 - \frac{\lambda} 4 x^4}$. Hence, the action and its first two derivatives are given by
\begin{equation}
\begin{gathered}
    S_\theta = \frac\alpha 2 x^2 + \frac\lambda 4 x^4
    \\
    \partial S_\theta = \alpha x + \lambda x^3
    \qquad
    \partial^2 S_\theta = \alpha + 3 \lambda x^2\text.
    \end{gathered}
\end{equation}
As a result we can write the score-matching loss function directly in terms of low-order expectation values, obtaining:
\begin{equation}
    \mathcal L(\alpha,\lambda)
    =
    (\alpha^2 - 6 \lambda) \langle x^2 \rangle
    + 2 \alpha\lambda \langle x^4\rangle
    + \lambda^2 \langle x^6 \rangle
    - 2 \alpha
    \text.
\end{equation}
Given the expectation values $\langle x^{2k}\rangle$ for $k=1,2,3$, this loss function is minimized at the true parameters $\alpha_*,\lambda_*$.

Now let us see how score matching connects to the Schwinger-Dyson relations discussed above for the distributions of the type \eqref{eq:generic_exp_family}.
At the minimum of the loss function $\mathcal L(\theta)$ defined in \eqref{eq:loss_estim}, derivatives with respect to $\theta$ must vanish:
\begin{equation}
\frac{\partial}{\partial \theta_i} \mathcal L(\theta) = 2 \langle (\partial Q_i) (\partial S)\rangle - 2 \langle \partial^2 Q_i\rangle = 0
\text.
\end{equation}
Defining $P_i = \partial Q_i$, this is precisely a Schwinger-Dyson relation of the form of Eq.~(\ref{eq:sd-1d}). This correspondence tells us that the minimum of the loss function automatically satisfies a certain subset of the Schwinger-Dyson equations\footnote{When performing score-matching we \emph{minimize} the score function, rather than merely seeking a saddle point. Just as the saddle-point condition corresponds to (some subset of) the Schwinger-Dyson relations, it may be seen that the minimization condition---the Hessian of $\mathcal L$ must be positive semi-definite---corresponds to (some subset of) positivity relations $\langle f^2 \rangle \ge 0$. This fact presumably has implications for the interplay of score-matching and sign problems, which we will not address in this work.}, determined by the free parameters $\theta$ that were used to define the ansatz $\pi(x;\theta)$.

We can also go in the other direction, showing that the score matching loss as a special case of Eq.~(\ref{eq:LA}). In order to recover the score loss, 
take the test polynomials to be $P_j = \partial Q_j$. Then $M_{ji} = \langle \partial Q_j \partial Q_i\rangle$, and taking $A= M^{-1}$, we recover the score-matching loss:
\begin{equation}
\begin{split}
    \mathcal{L}^{\rm SM}(\theta) &= \sum_{i,j}r^T_{i}(\theta)M^{-1}_{ij}r_j(\theta)\\
    &=
    b^T M b - 2 b^T \theta + \theta^T M \theta
    \;.
    \end{split}
\end{equation}
To summarize, we showed that score matching loss that has an interpretation of matching of score functions in \eqref{eq:score_matching_loss}, is a special case of our family of estimators defined in Eq.~(\ref{eq:LA}). These generalized family can be constructed starting with a sufficient number of Schwinger-Dyson relations that need to be satisfied at the minimum of the loss function. With this principle in mind, in the next Section we show how the general construction is essential for dealing with Grassmann variables that are used to model fermions.

\section{Grassmann variables}\label{sec:grassmann}

Path integrals for fermionic theories are commonly constructed as integrals over Grassmann variables. Functions of Grassmann variables are not maps from one space to another, but instead formal expressions, with rules for linear combination, multiplication, differentiation, and integration. For completeness, we include a brief review of the properties of Grassmann variables, and path integrals built off of them, in Appendix~\ref{app:grassmann}. We refer the reader to a standard reference (e.g.~\cite{Peskin:1995ev}) for more details.

A lattice path integral for a fermionic system typically has the form
\begin{equation}
    Z = \int \left(\prod_i d \bar\eta_i \,d\eta_i\right)
    e^{-S(\bar\eta,\eta)}
    \text.
\end{equation}
Here $\bar\eta_i$ and $\eta_i$ are independent Grassmann variables on lattice site $i$, which are by convention taken to transform into each other under complex conjugation. Needless to say, this path integral does not directly correspond to any statistical system in the usual sense. The action $S(\bar\eta,\eta)$ is not describing the logarithm of a probability distribution on any space. Instead it is to be treated as a formal expression. Nevertheless, expectation values are defined in the usual way, and we can ask to determine properties of $S(\bar\eta,\eta)$ given those expectation values.

\subsection{Generalities}
To guide the eye, we will again work with a concrete example before giving the general procedure. We consider a one-site model, in which the action is necessarily quadratic in the Grassmann degrees of freedom. The partition function of this model is given by
\begin{equation}
	Z_1(\theta^{*}) = \int d\bar\eta d \eta\, e^{-\theta^{*} \bar\eta \eta}
	\text,
\end{equation}
and is readily evaluated to be $Z_1(\theta^{*}) = \theta^{*}$. Expectation values in this system are defined in the usual way. A straightforward application of the rules of Grassmann arithmetic shows that the only nontrivial expectation value is $\langle \eta \bar\eta\rangle = (\theta^{*})^{-1}$.

This explicit evaluation of the expectation value immediately translates into a learning algorithm, but we can get the same result in a more generalizable way by considering Schwinger-Dyson relations. The Schwinger-Dyson relations are derived in the same way as in the bosonic case, and for this one-site model there are two:
\begin{equation}
\begin{split}
    \Big\langle\frac{\partial}{\partial \eta} \eta \Big\rangle &= 
    -\Big\langle\eta\frac{\partial}{\partial \eta} \theta^{*} \bar\eta\eta \Big\rangle
    \text{, and}
    \\
    \Big\langle\frac{\partial}{\partial \bar \eta} \bar\eta \Big\rangle &= 
    -\Big\langle\bar\eta\frac{\partial}{\partial \bar\eta} \theta^{*} \bar\eta\eta \Big\rangle
    \text.
    \end{split}
\end{equation}
Thanks to the symmetry $\bar\eta \leftrightarrow \eta$, these two expressions are identical, and we recover $1 = \theta^{*} \langle \eta\bar\eta \rangle$ as expected above. Note that the Schwinger-Dyson equations which would be derived by considering (for example) $\int d\bar\eta d\eta\, \frac{\partial}{\partial \eta}(\bar\eta e^{-S})$ trivially vanish.

At this point it is straightforward to construct a loss function by means of Eq.~(\ref{eq:LA}). The Schwinger-Dyson relations above define the residual vector $r$, and we may take $A$ to be the identity. The following loss function is thus obtained:
\begin{equation}
    \mathcal L(\theta)
    = \left(1 - \theta\langle \bar\eta\eta\rangle\right)^2\text.
\end{equation}
We may now generalize to the case of arbitrarily many Grassmann fields. The general Schwinger-Dyson relation has the form
\begin{equation}
    \Big\langle \partial_{\eta/\bar{\eta}}P \Big\rangle
    =
    -\Big\langle P \partial_{\eta/\bar{\eta}} S\Big\rangle
    \text,
\end{equation}
for Grassmann-odd polynomials $P$. For a sufficiently large basis of such relations and expectation values, one can recover the parameters of the action. The general rules and conditions for identifiability mirror those mentioned in Section~\ref{sec:two}.

Notice that score matching as defined in \eqref{eq:score_matching_loss} does not apply to Grassmann theories in an unambiguous way, because a suitable norm must be defined on Grassmann variables. In the one-dimensional example above, a sensible score-matching loss function is
\begin{equation}\label{eq:grassman_sf}
    \mathcal L(\theta)
    = \Big\langle
    \Big\Vert\frac{\partial S(\theta_*)}{\partial \eta}
    -
    \frac{\partial S(\theta)}{\partial \eta}
    \Big\Vert^2
    \Big\rangle
    \text.
\end{equation}
Here the norm $\Vert f \Vert^2$ is defined as $\bar f f$, where Grassmann variables are transformed under complex conjugation in the conventional way as $\bar\eta$. Direct evaluation of this loss function reveals that it has a unique minimum at $\theta=\theta_*$ as desired. However we can already see that this choice is ``brittle'': if instead we had taken the derivatives with respect to $\bar\eta$, then the expectation value would be negative $\mathcal L(\theta)$ would be unbounded below.

Once again, this loss cannot be estimated just from observations as it explicitly depends on the unknown parameter, $\theta^{*}$, that we are trying to learn. As in Section~\ref{sec:two}, we must remove the explicit dependence of $\mathcal L(\theta)$ on $\theta^{*}$, dropping all $\theta$-independent terms. The result is
\begin{equation}\label{eq:sm-grassmann-loss}
    \mathcal L(\theta) =
    \Big\langle \frac{\partial S_\theta}{\partial \bar\eta} \frac{\partial S_\theta}{\partial \eta}\Big\rangle
    + 2 \Big\langle \frac{\partial}{\partial \eta}\frac{\partial}{\partial \bar\eta} S_\theta\Big\rangle
    + \mathrm{const}\text.
\end{equation}
This loss function, a naive translation of the score-matching loss to the Grassmann case, is usable only in the special case where the action $S$ is quadratic in the Grassmann variables. Higher-order terms in the action vanish when the norm is taken (because the norm defined above obeys $||\bar\eta \eta|| = 0$).

\subsection{Lattice Thirring model}
\begin{figure}
\centering\includegraphics[width=0.9\linewidth]{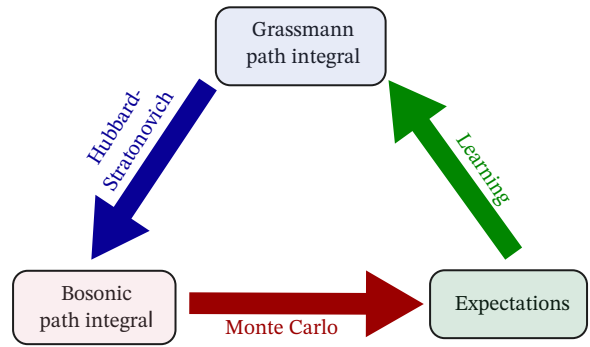}
	\caption{Workflow for the demonstration of learning of Grassmannian actions in the case of the lattice Thirring model. We begin with an action written purely in terms of Grassmann fields. Bosonization (via Hubbard-Stratonovich) yields a theory in which the fermions can be integrated out, leaving only c-number fields. Expectation values in this theory are obtained by Monte Carlo, and from those expectation values we use Schwinger-Dyson-based learning to recover directly the original (purely Grassmann) theory.\label{fig:triangle}}
\end{figure}
To demonstrate learning in a theory with Grassmann fields, we consider a lattice Thirring model defined with staggered fermions. This is a two-dimensional statistical theory with a single Grassmann variable $\chi$ at each site. For any polynomial $\mathcal O$ of the Grassmann fields, an expectation value is defined by the Grassmann path integral
\begin{equation}
	\langle \mathcal O \rangle =
	\frac
	{\int d\bar\chi d\chi\,\mathcal O(\bar\chi,\chi) e^{-S_{\mathrm{Th}}(\bar\chi,\chi)}}
	{\int d\bar\chi d\chi\, e^{-S_{\mathrm{Th}}(\bar\chi,\chi)}}
	\text.
\end{equation}
The action $S_{\mathrm{Th}}$ consists of the standard action of free staggered fermions~\cite{Kogut:1974ag}, combined with a quartic vector interaction:
\begin{equation}
    S_{\mathrm{Th}} =
	m \sum_x \bar\chi_{x}\chi_{x}
	+
	\sum_{\langle x,x'\rangle}\bar{\chi}_x M_{xx'}\chi_{x'}
    +
    \frac{g^2}{2}\sum_{x,\nu}j_{\nu}^2(x)    \text.
    \label{eq:S_Th}
\end{equation}
Here the second sum is taken over all pairs of neighboring lattice sites on an $L\times L$ lattice with periodic boundary conditions, and $\nu = 0,1$ denotes the Euclidean directions. The massless staggered Dirac operator is 
\begin{equation}\begin{split}
    M_{xx'} &= \sum_{\nu = 0}^1\frac{\eta_{x,\nu}}{2}\left(\delta_{x+\hat{\nu},x'}-\delta_{x-\hat{\nu},x'}\right)\\
    \eta_{x,0} &= 1,\quad\eta_{x,1} = (-1)^{x_0}\;\text,
    \end{split}
    \label{eq:masslessDirac}
\end{equation}
while the conserved staggered vector current on the link $(x,\nu)$ is
\begin{align}
    j_{\nu}(x) &= \frac{\eta_{x,\nu}}{2}(\bar\chi_x \chi_{x+\hat\nu}
      + \bar\chi_{x+\hat\nu}\chi_x)\;.
    \label{eq:current} 
\end{align}
In the continuum limit, this results in a theory of two flavors of (two-component) Dirac fermions. For our purposes we will not need to consider the continuum limit: we are only interested in the lattice theory on its own.

In order to numerically simulate the Thirring model (via lattice Monte Carlo), we introduce a bosonic auxiliary field in such a way that the action is quadratic in the fermion fields. This allows the Grassmann variables to be integrated out analytically. Using the Hubbard-Stratonovich formalism, we have for each link $(x,\nu)$,
\begin{align}
    \exp\left(-\frac{g^2}{2}j_{\nu}\right) \propto  \int dA_\nu\exp\Big[-\frac{A_\nu^2}{2} - i gA_\nu j_\nu\Big]\;.
\end{align}
The resulting action, after integrating out the fermions, depends only on the auxiliary vector field $A$:
\begin{equation}
S_{\mathrm{Th,bose}}
= -\log\det D[A,g]
  + \frac{1}{2}\sum_{x,\nu}
    A^2_\nu(x).
    \label{eq:STh_bose}
\end{equation}
The staggered Dirac operator in the $A$--background is
\begin{align}
D_{xx'}[A,g] &= m\,\delta_{x,x'} + \sum_{\nu=0}^1
    \frac{\eta_{x,\nu}}{2}
    \biggl[
      \bigl(1 + igA_\nu(x)\bigr)
      \delta_{x+\hat{\nu},x'}     \notag\\
&\qquad\qquad\qquad
      - \bigl(1 - igA_{\nu}(x)\bigr)
        \delta_{x-\hat{\nu},x'}
    \biggr].
    \label{eq:staggeredDirac_g}
\end{align}

As detailed in Appendix~\ref{sec:SD_Th}, the Schwinger-Dyson relations used are those defined by taking $P = \partial_{\bar{\chi}_k}Q$ in the equation:
\begin{align}
    \langle \partial_{\chi_k}P\rangle = -\langle P\partial_{\chi_k}S\rangle\;, 
\end{align}
for $Q \in \{Q^{(1)},Q^{(2)}\}$, where $Q^{(1)}_{ab} = \bar\chi_{a}\chi_{b}$ and $Q^{(2)}_{a\rho} = \bar{\chi}_{a+\rho}\chi_{a+\rho}\bar{\chi}_{a}\chi_{a}$. 

To demonstrate Schwinger-Dyson learning of Grassmann-valued field theories, we perform the following procedure, portrayed in Figure~\ref{fig:triangle}. We first choose bare couplings $m$ and $g$. Using $S_{\mathrm{Th,bose}}$, we sample field configurations $A$. From these configurations we compute the fermion propagator, and from there, a set of observables sufficient to infer the bare parameters $m$ and $g$. Finally, we fit the Schwinger-Dyson relations obtained from $S_{\mathrm{Th}}$ into the loss in Eq.~\eqref{eq:LA} after setting the matrix $A = \mathbb{I}$, obtaining estimated bare couplings $\tilde m$ and $\tilde g$. This style of demonstration is selected in part to show that the Schwinger-Dyson learning procedure operates directly on (formal) expectation values, rather than requiring samples from a probability distribution.

Note that the advantage of the developed procedure is that unlike in \cite{Shukla:2025wze},
we are able to directly learn the original action which is defined on Grassmann-valued fields, although we cannot sample from this action because it does not define a probability distribution on any space. The two systems are related by an exact duality: all expectation values are the same on both sides of the Hubbard-Stratonovich transformation.

\begin{figure}
	\centering
	\includegraphics[width=0.95\linewidth]{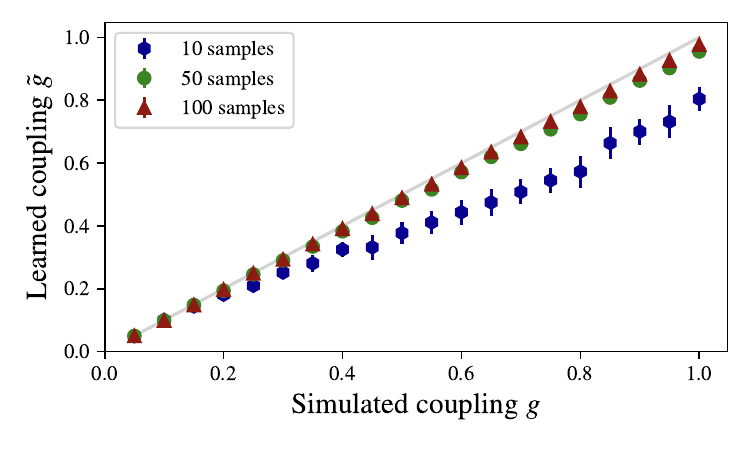}
    \caption{Estimates of the bare coupling of the Thirring model, as the bare coupling used for simulation is changed. All calculations are performed at a bare mass of $m = 0.1$ on a $12 \times 12$ lattice.\label{fig:thirring}}
\end{figure}

The results of this learning procedure are shown in Figure~\ref{fig:thirring}. We examine multiple values of the coupling $g$, holding the mass $m=0.1$ and the lattice size $L=12$ fixed. All parameters are reported in lattice units $a=1$. For each value of bare coupling $g$, two fits are performed, one using $50$ samples and the other using $100$. Errors shown are obtained with the statistical bootstrap, and represent one standard deviation of the spread of couplings reported by the learning algorithm.

Figure~\ref{fig:thirring} makes apparent two facts about the learning algorithm. First, quite a small number of samples is sufficient to obtain reasonably high precision (at the strongest couplings, a few percent) in the determination of the bare coupling. Second, the estimator defined by the learning algorithm is \emph{biased} for finite samples (recovering the correct parameters in the limit of the large number of samples). We note that the presence of a bias is standard and explicitly appears in the analysis of nearly all learning algorithms \cite{lokhov2018optimal}.  

\section{Gauge symmetries}\label{sec:gauge}
In this section we consider one final extension of the learning formalism: the case of theories with gauge symmetries. From the point of view of a Euclidean path integral (or equivalently, a statistical field theory), a gauge symmetry is a local symmetry of the action, a family of automorphisms $\phi$ of field space such that $S = S \circ \phi$. Gauge symmetries occur in the lattice field theories of greatest interest for the standard model, and in particular in lattice QCD.

From one perspective, gauge symmetries introduce no complications whatsoever. A theory that happens to have gauge symmetry can be thought of as just a particular point in a large family of theories that generically lack gauge symmetry, and the formalism of Section~\ref{sec:two} applies unmodified. However, in the context of gauge theories one ordinarily applies an additional constraint, which is that no object which is not gauge-invariant is defined\footnote{When considering gauge theories in the Hamiltonian formalism, this is accomplished by declaring the Hilbert space of the theory to be those states which are not altered by gauge transformations. From this perspective there is no such thing as ``gauge symmetry'', but instead a restriction of the space of states and operators.}. It is therefore important\ that, both in the case of score-matching and Schwinger-Dyson-based learning, only gauge-invariant expectation values are considered\footnote{The restriction to gauge-invariant quantities is not just important for philosophical reasons, but because there are a variety of numerical methods that give access only to gauge-invariant quantities---and this is true of \emph{all} experimental methods.}. The first subsection below shows how this is accomplished, and the second subsection applies this formalism, together with the results on Grassmann-valued fields of the previous section, to detail the learning problem of lattice QCD. 

\subsection{Generalities}
Yang-Mills (in $4$ spacetime dimensions) is the quintessential example of a gauge theory, and a natural starting point for studying QCD. In this section we consider the Wilson discretization of $SU(N)$ Yang-Mills~\cite{Wilson:1974sk}.

This theory is defined on a hypercubic lattice in four dimensions. Define $V$ to be the total number of lattice sites. To each edge in the lattice (of which there are $4V$) is associated a single $SU(N)$ degree of freedom. Thus a field configuration $U$ consists of $4V$ $SU(N)$ matrices. We consider $U$ to be a spacetime-indexed function of the lattice: $U_\mu(x)$ lives on the link between $x$ and $x + \hat \mu$. 

On this space of field configurations we define the Wilson action as follows:
\begin{equation}
	S_{\mathrm{YM}}(U) = \frac{2}{g^2} \sum_{x,\mu<\nu}\Re \Tr \left(I - P_{\mu\nu}(x)\right)
    \text.
\end{equation}
Here $g$ is a coupling constant, and the only free parameter in the action. The sum is taken over all ``plaquettes''---that is, over all loops of $4$ distinct links. The notation $P_{\mu\nu}(x)$ refers to the product of links along that loop:
\begin{equation}
    P_{\mu\nu}(x) = U_\nu(x)^\dagger U_\mu(x+\hat\nu)^\dagger U_\nu(x+\hat \mu) U_\mu(x)
    \text.
\end{equation}
With this action written, it is a small matter to see the gauge symmetry. For each site $x$, select a matrix $V(x)$. The matrices $U$ then transform according to
\begin{equation}
    U_\mu(x) \mapsto V(x+\hat\mu) U_\mu(x) V^\dagger(x)
\end{equation}
which is manifestly a symmetry of $S_{\mathrm{YM}}$.

Although it is not directly visible from the Yang-Mills action, one can imagine that at each site $x$ lives a vector space $V^{(x)} \approx \mathbb C^{N}$. One should think of $U_\mu(x)$ not as an automorphism of $\mathbb C^N$, but instead as a linear map $V^{(x)} \rightarrow V^{(x + \hat \mu)}$. The gauge transformations at $x$ correspond to automorphisms of $V^{(x)}$ (acting trivially on the spaces that live at all other sites), and this motivates the transformation law of the fields $U$ given above. This view also makes it clear that if $U$ is the link from $x\rightarrow y$, then $U^\dagger$ serves a symmetric role as the link from $y\rightarrow x$.

Expectation values are defined in the usual way, via a path integral over the space $SU(N)^{\otimes 4 V}$ of field configurations
\begin{equation}
    \langle \mathcal O \rangle
    = \frac{\int\left[\prod_{\mu,x}dU_\mu(x)\right]\mathcal{O}(U)\,\mathrm{e}^{-S_{\rm YM}[U]}}{\int \left[\prod_{\mu,x}dU_\mu(x)\right] \mathrm{e}^{-S_{\mathrm YM}[U]}}\text.
\end{equation}

To set up the learning problem, we imagine that we are given access to expectation values over an empirical distribution which is drawn from the family of actions defined by
\begin{equation}
    S_\beta(U) = \beta \sum_{x,\mu<\nu}\Re \Tr \left(I - P_{\mu\nu}(x)\right)
    \text.
\end{equation}
By assumption there exists some $\beta^{*}$ such that the empirical distribution is proportional to $e^{-S_{\beta^{*}}(U)}$. We wish to learn this value of $\beta^{*}$ by inspection of the expectation values.

To illustrate the construction, we first consider the score matching approach, which provides intuition for gauge-invariant observables needed. We then show how the parameters may also be determined from Schwinger-Dyson relations. Let us begin by defining the score function. As discussed, we impose that this score function be gauge-invariant. A natural choice is:
\begin{equation}\label{eq:ym-sf}
    \mathcal L(\beta)
    =
    \Big\langle
    \left\Vert
    \frac{\partial}{\partial U_{\mu}^a(x)}\log p(U) - \frac{\partial}{\partial U_{\mu}^a(x)}\log \pi(U;\beta)
    \right\Vert^2
    \Big\rangle
    \text.
\end{equation}
As before, the derivative operator is defined as a vector indexed over spacetime ($x$), direction of links ($\mu$), and color ($a$). Here the partial derivative is defined as follows. For any function $f(U)$, we define
\begin{equation}
    \frac{\partial}{\partial U^a} f(U)
    =
    \lim_{\epsilon \rightarrow 0} \frac{f(e^{i \epsilon T_a} U) - f(U)}{\epsilon}
    \text.
\end{equation}
Here the $T_a$ are a basis of generators of $SU(N)$. When $N=2$ these would conventionally be the Pauli matrices; for $N=3$ they are the eight Gell-Mann matrices. In any case we adopt the normalization given by the orthonormality relation $\Tr T_i T_j = 2 \delta_{ij}$\text.

To see that Eq.~(\ref{eq:ym-sf}) is gauge-invariant, we must consider how these derivatives behave under gauge transformations. Consider performing a gauge transformation by $V \in SU(N)$ at site $x$. Then the derivative with respect to $U(x)$ transforms according to
\begin{equation}\label{eq:dplaquette}
    \frac{\partial}{\partial U^a(x)} P_{\mu\nu}
    \mapsto
    \rho_{\mathrm{adj}}(V)_{ab} \frac{\partial}{\partial U^b(x)} P_{\mu\nu}
    \text.
\end{equation}
where $\rho_{\mathrm{adj}}$ is the adjoint representation (written in the same basis as was chosen to define $T_a$) of the group $SU(N)$.

The $L_2$ norm used to define $\mathcal L$ is now gauge-invariant due to the fact that a representation is a homomorphism and the group is unitary: $\rho(V)^\dagger \rho(V) = \rho(V^\dagger V) = 1$. By similar reasoning we find that the diagonal of the Hessian is also gauge-invariant.

We now proceed to remove the explicit $\beta^{*}$ dependence from $\mathcal L$ in the usual way, neglecting terms which are independent of $\beta$ and integrating by parts. Up to a constant, the score-matching loss is now
\begin{equation}
    \mathcal L(\beta)
    = \Big\langle \Big| \frac{\partial}{\partial U^a_\mu(x)} S \Big|^2\Big\rangle - 2\Big\langle \frac{\partial^2}{\partial U^a_\mu(x)^2} S \Big\rangle\text,
\end{equation}
which is manifestly gauge-invariant as discussed. The minimum of this score function satisfies 
\begin{equation}\label{eq:dL_YM}
    0 = \nabla \mathcal L
    = \beta \Big\langle \Big| \frac{\partial}{\partial U^a_\mu(x)}  R \Big|^2\Big\rangle - \Big\langle \frac{\partial^2}{\partial U^a_\mu(x)^2} R \Big\rangle
\end{equation}
where we have defined
\begin{equation}
    R =  \sum_{x,\mu<\nu}\Re \Tr \left(I - P_{\mu\nu}(x)\right)\text.
\end{equation}
The second derivative term is given by the quadratic Casimir of the $SU(N)$ fundamental representation:
\begin{equation}
    C_2(N) I  = \sum_i T_i^2 = 2 \left(\frac{N^2-1}{N}\right) I\text.
\end{equation}
The other term---the norm of the first derivative---is evaluated by using the Fierz identity
\begin{equation}
    \sum_a T_a^{ij} T_a^{kl}
    =
    2 \left(
    \delta^{il}\delta_{jk}
    -
    \frac 1 N
    \delta^{ij}\delta^{kl}
    \right)
    \text.
\end{equation}
The second-derivative term only affects one plaquette at a time, and becomes a sum over all plaquettes. The first-derivative term is more complicated, since the same $\frac{\partial}{\partial U^a_\mu(x)}$ derivative acts nontrivially on any plaquette which includes the link from $x$ to $x+\hat \mu$. The result is a sum over the $36$ (ordered) pairs of plaquettes which involve that link:
\begin{widetext}
\begin{equation}
    \left|\frac{\partial}{\partial U^a_\mu(x)} R\right|^2
    = \sum_{P_1,P_2}\Big[\text{Re}\,\text{Tr}(P_1 P_2) + \text{Re}\,\text{Tr}(P_1^\dagger P_2) - \frac{2}{N}\text{Re}\,\text{Tr}(P_1) \text{Re}\,\text{Tr}(P_2)\Big]\text.
\end{equation}
Here $P_1$ and $P_2$ are products of four links, beginning and ending at site $x$, corresponding to plaquettes which include the link between $x$ and $x+\hat\mu$.

Putting this all together, and making use of the lattice symmetries of translation and rotation invariance, we find that $\beta^{*}$ can determined by two expectation values:
\begin{equation}\label{eq:ym-sm-final}
    \beta^{*}
    \sum_{P_1,P_2}\left<\Big[\text{Re}\,\text{Tr}(P_1 P_2) + \text{Re}\,\text{Tr}(P_1^\dagger P_2) - \frac{2}{N}\text{Re}\,\text{Tr}(P_1) \text{Re}\,\text{Tr}(P_2)\Big]\right>
    =
    2 \frac{N^2-1}{N} \left<\Re \Tr P_{\mu\nu}(x)\right>
    \text.
\end{equation}
\end{widetext}

This has the form of a Schwinger-Dyson relation, but as before, this is not the only Schwinger-Dyson relation we could have used to determine the value of $\beta$. The most general Schwinger-Dyson relation on the Yang-Mills lattice reads
\begin{equation}
    \Big\langle \frac{\partial P}{\partial U^a(x)}\Big\rangle = \Big\langle P\frac{\partial S}{\partial U^a(x)}\Big\rangle
    \text,
\end{equation}
for some choice of site $x$ and adjoint index $a$. Most such relations will involve, and therefore determine, $\beta^{*}$. 
However, this equation does not in general relate two gauge-invariant quantities. Suppose for example that $P$ is the trace of some plaquette---then neither its derivative, nor its product with the derivative of the action, is gauge-invariant.

To ensure gauge-invariance, it is necessary that the function $P$ be chosen to transform in the adjoint representation of the site at which the derivative is being taken, and then sum over all $N^2-1$ components. That is, fixing a lattice site $x$, the relation
\begin{equation}\label{eq:sd_U}
    \sum_a\Big\langle \frac{\partial \Tr T_a W(x)}{\partial U^a(x)} \Big\rangle = \sum_a\Big\langle \Tr T_a W(x)\frac{\partial S}{\partial U^a(x)} \Big\rangle
\end{equation}
is manifestly gauge-invariant for any loop $W(x)$ begins and ends at site $x$. The simplest example is of course to take $W$ to be a plaquette, as discussed above.
\subsection{Lattice Quantum Chromodynamics}
We conclude by discussing QCD on the lattice, which requires all the technology developed in this paper so far. For concreteness we consider a lattice theory of gauge fields using the Wilson action, and Wilson-discretized fermions (see~\cite{Gattringer2010,Montvay1994} for standard references on lattice QCD). The action of this theory is a function both of the gauge fields and the (Grassmann-valued) fermions:
\begin{equation}\label{eq:s_QCD}
    S_{\mathrm{QCD}}(U,\bar\psi,\psi) = 
    S_{\mathrm{YM}}(U) + \bar\psi^{(f)}(x) D(U) \psi^{(f)}(y)
    \text,
\end{equation}
where $D(U)$ is the lattice Dirac operator. The quark fields $\psi^{(f)}(x)$, for each flavor $f$ and at each site $x$, are $4 N$ component objects, implicitly indexed by color and Dirac index. For Wilson fermions we take the Dirac matrix $D$ to be
\begin{widetext}
\begin{equation}
    D_{\mathrm W}(U)_{x\alpha a}^{y \beta b}
    = \left( m_f + 4 r \right)\delta_{xy}\delta_{\alpha\beta}\delta_{ab} - \frac{1}{2}\sum_{\mu}\bigg[\left[r-\gamma_\mu\right]_{\alpha\beta} \delta_{y,x+\hat{\mu}} \left[U_\mu(x)\right]_{ab}
    +\left[r+\gamma_\mu\right]_{\alpha\beta} \delta_{y,x-\hat{\mu}} \left[U^\dagger_\mu(x-\hat{\mu})\right]_{ab}  \bigg]
    \text,
\end{equation}
\end{widetext}
where $r$ is the Wilson parameter (typically $r=1$), removing fermion doublers at the cost of breaking the chiral symmetry explicitly. Latin indices ($a,b$) refer to color, and Greek ($\alpha,\beta$) to spin components. The parameters of the action are $\beta$ in $S_{\mathrm{YM}}$ and masses $m_f$ of quarks, which we collectively denote as $\theta$ in the remaining discussion.

As in the case of Yang-Mills, it is straightforward to construct a manifestly gauge-invariant score function via the $L_2$ norm:
\begin{align}\label{eq:qcd-sf}
    \mathcal L(\theta)
    &= \sum_{f}\Big\langle
    \left\Vert
    \frac{\partial}{\partial \psi^{(f)}}\Big[ S_{\mathrm{QCD}}- S^{(\theta)}_{\mathrm{QCD}}\Big] \right\Vert^2
    \Big\rangle\nonumber\\
    &+
    \Big\langle
    \left\Vert
    \frac{\partial}{\partial U^a_{\hat\mu}(x)} \Big[ S_{\mathrm{QCD}} - S^{(\theta)}_{\mathrm{QCD}}\Big]
    \right\Vert^2
    \Big\rangle
    \text.
\end{align}
Here the norm, defined in Eq.~(\ref{eq:grassman_sf}), is taken in the $N$-dimension color space, spacetime, and Dirac space for the first term. The norm in the second term follows the definition in Eq.~(\ref{eq:ym-sf}).
Gauge invariance of the score function follows from the transformation rules of the derivatives of the action. For typical quark bilinear terms in the action,
\begin{equation}
    B(x,\mu) = \bar\psi(x) U_{\mu}(x) \psi(x+\mu)
    \text,
\end{equation}
their derivatives with respect to quark and gauge fields transform in the (anti-)fundamental and adjoint representation respectively: 
\begin{align}
    &\frac{\partial}{\partial U^a_{\mu}(x)} B(x,\mu)
    \mapsto   \rho_{\mathrm{adj}}(V(x))_{ab} \frac{\partial}{\partial U^b_{\mu}(x)} B(x,\mu) \\
    &\frac{\partial}{\partial \bar\psi_{a}(x)} B(x,\mu) 
    \mapsto  \rho_{\mathrm{fund}}(V(x))_{ab} \frac{\partial}{\partial \bar\psi_{b}(x)} B(x,\mu) 
    \text.
\end{align}
Note that we have dropped irrelevant Dirac and flavor indices on quark fields. The derivative of the action with the quark fields $\psi$ transforms in the fundamental representation, making the first term in Eq.~(\ref{eq:qcd-sf}) gauge-invariant. Combining the transformation rule of the plaquettes, Eq.~(\ref{eq:dplaquette}), and of quark terms $Q(x,\hat\mu)$ above, the derivatives of the action with gauge fields transform under the adjoint representation, thus the second term in the score function is also gauge-invariant. 

After dropping constant terms from the empirical distribution, the score function reads 
\begin{align}
    \mathcal{L}(\theta) &= \sum_{f} \Big\langle \Big| \frac{\partial}{\partial \psi^{(f)}} S^{(\theta)}_{\mathrm{QCD}} \Big|^2 + 2\frac{\partial}{\partial \psi^{(f)}}\frac{\partial}{\partial \bar\psi^{(f)}} S^{(\theta)}_{\mathrm{QCD}} \Big\rangle \nonumber\\
    &+ \Big\langle \Big| \frac{\partial}{\partial U^a_\mu(x)} S^{(\theta)}_{\mathrm{QCD}} \Big|^2 - 2\frac{\partial^2}{\partial U^a_\mu(x)^2} S^{(\theta)}_{\mathrm{QCD}}\Big\rangle \;\text.
\end{align} 

For completeness we briefly discuss the construction of gauge-invariant Schwinger-Dyson equations in lattice QCD. Schwinger-Dyson equations for lattice QCD can be constructed via the derivatives of functions of $U,\bar\psi,\psi$ with respect to these fields. When differentiating with respect to the gauge fields we again obtain relations of the form of Eq.~(\ref{eq:sd_U}); when differentiating with respect to quark fields we obtain:
\begin{equation}\label{eq:sd_quark}
    \Big\langle \sum_a\frac{\partial P_a}{\partial \bar\psi^{(f)}_{\alpha,a}(x)}\Big\rangle = -\Big\langle \sum_a P_a\frac{\partial S}{\partial \bar\psi^{(f)}_{\alpha,a}(x)}\Big\rangle
    \text.
\end{equation}
To have the both sides of the equation be manifestly gauge-invariant, we want to choose as $P_a$ a function that transforms in the anti-fundamental representation. One simple choice is $P_{\alpha',a}=\bar\psi^{(f)}_{\alpha',a}(x')$, for which the Schwinger-Dyson equation reads
\begin{equation}
    N \delta_{\alpha\alpha'} \delta_{xx'} = -\sum_{y,\beta,a,b} \Big\langle \bar\psi^{(f)}_{\alpha',a}(x)  D_{\mathrm W}(U)_{x\alpha a}^{y \beta b} \psi^{(f)}_{\beta,b}(y) \Big\rangle\;\text.
\end{equation}
This Schwinger-Dyson relation allows the quark mass $m_f$ to be determined; the gauge coupling $\beta$ is not involved and cannot be determined from this equation. A Schwinger-Dyson equation that fixes $\beta$ can be derived via Eq.~(\ref{eq:sd_U}).

\section{Discussion}\label{sec:discussion}

We have presented a broad family of learning methods based on the lattice Schwinger-Dyson relations. Because Schwinger-Dyson relations are available for Grassmann-valued fields, our method allows learning of couplings in theories with fermions. Similarly, as (a certain subset of) lattice Schwinger-Dyson equations are gauge-invariant, we obtain a learning algorithm for lattice gauge theories (including QCD) which operates exclusively on gauge-invariant data. Score matching, a venerable approach to learning exponential families of distributions, is a specific case of a broad family of learning methods based on Schwinger-Dyson relations. However, by expanding the family of learning methods considerably beyond score matching, we obtain a way to learn the action parameters with potentially lower cost, as there is more choice in the expectation values that must be measured. A concrete case where the Schwinger-Dyson-based learning procedure is advantageous, in the sense of providing an estimator of lower variance than that of score matching, is presented in Appendix~\ref{app:not-optimal}.

Throughout this work, we have neglected the inevitable statistical errors in estimates of expectation values, and we have neglected the computational costs associated to determining those expectation values. The existence of these errors and costs is one of the motivations for our generalization to arbitrary Schwinger-Dyson relations. As shown in the appendix, in general the Schwinger-Dyson relations represented in the score function are not the ones with that yield the smallest statistical errors. We leave to future work a proper treatment of these issues, which are critical to applying these methods on larger lattices.

This work is motivated in part by the application suggested in~\cite{Shanahan:2018vcv,Shukla:2025wze}, of studying renormalization flows on the lattice. We anticipate that the technology developed here will allow that renormalization program to be applied efficiently to (among others) theories of interacting fermions.

This paper has been entirely limited to theories defined via Euclidean lattice path integrals (whether bosonic or Grassmannian). Schwinger-Dyson equations are also available for, and can in principle be used to learn from, correlation functions computed from a real-time path integral. Whether such a method can be of use is unclear. Real-time lattice correlators are famously difficult to extract due to the sign problem~\cite{Lawrence:2021izu}, even once a sensible (i.e.~unitary) action has been defined~\cite{Kanwar:2021tkd}. Similarly, we have not addressed learning from Hamiltonian data, although prior work has studied this in both the ground state~\cite{Qi:2019brg} and at finite temperature~\cite{Bakshi:2023csv}.

\subsection{Remark on integration contours}

One should not make the mistake of thinking that Schwinger-Dyson relations determine all facts about a (lattice) quantum field theory. Crucial nonperturbative physics, such as instantons, are not constrained by these equations. A simple example is accessible analytically: consider a one-site lattice model, with expectation values given by
\begin{equation}
    \langle f \rangle := \frac{\int_\gamma dz\, f(z) e^{-z^4}}{\int_\gamma dz\, e^{-z^4}}
    \text.
\end{equation}
Here the integrals are to be interpreted as complex contour integrals, over some contour $\gamma \subset \mathbb C$. There are multiple contours $\gamma$ that might be chosen. For instance one can take $\gamma = \mathbb R$, giving the ``obvious'' integral along the real line. The expectation values are also convergent, for polynomial $f$, when we select $\gamma = i \mathbb R$---here the integral is performed along the imaginary axis. The resulting expectation values are not equal; however, both sets of expectation values satisfy the Schwinger-Dyson relations!

One can see the same phenomenon by considering the Schwinger-Dyson relations in isolation, without reference to the notion of an integration contour. A complete set of Schwinger-Dyson relations for this simple model reads, for positive integers $n$: $\langle z^{n-1}\rangle = 4 \langle z^{n+3}\rangle$.
These are recursion relations, and there exists a solution for any choice of $\langle z^2\rangle$. This choice may be seen to be degenerate with the choice of homology class (albeit over $\mathbb R$, instead of $\mathbb Z$ as usual).

Knowledge of a sufficient number of expectation values is sufficient in principle to fix the choice of contour---or equivalently, to select a single solution from vector space of solutions to the Schwinger-Dyson relations. However the methods in this paper do not provide a way to extract this information efficiently.

This question regarding what contour should be used is relevant to physical theories. The practical effect of choosing the wrong integration contour is either to neglect nonperturbative ``instanton-like'' contributions~\cite{Lawrence:2023woz}, or to render the integral entirely non-convergent (as in the Euclidean gravitational path integral~\cite{Gibbons:1978ac}, where the question of the correct choice of integration contour is indeed central~\cite{Kontsevich:2021dmb,Witten:2021nzp,Banihashemi:2024lzcv}). In quantum field theory, the contour ambiguity may in principle be fixed by demanding unitarity (in Euclidean quantum field theory, reflection positivity); in general it is unclear how one should select the ``physical'' integration contour.

\section*{Acknowledgments}
S.S., A.J.~and A.Y.L.~acknowledge support from the U.S.~Department of Energy/Office of Science Advanced Scientific Computing Research Program. S.L.~is supported by a Richard P.~Feynman fellowship from the Laboratory Directed Research and Development (LDRD) program of Los Alamos National Laboratory (LANL). Y.Y.~is supported by a Darleane C.~Hoffman fellowship from the LANL LDRD program. Los Alamos National Laboratory is operated by Triad National Security, LLC, for the National Nuclear Security Administration of U.S.~Department of Energy, under Contract No.~89233218CNA000001.

\appendix
\section{Estimator variances in one dimension}\label{app:not-optimal}

The purpose of this appendix is to demonstrate that the coupling estimators yielded by score matching are not, in general, most efficient estimators available, even within the family of estimators constructed from Schwinger-Dyson equations.

There are multiple meanings that can be assigned to the phrase ``most efficient'', and which one is relevant depends on context. For example in lattice QCD, a premium is placed on minimizing the number of inversions of the Dirac matrix that are required. Here we show that Schwinger-Dyson learning can yield estimators of smaller variance (and therefore, requiring fewer samples for a fixed precision) than score matching.

Consider the exponential family of distributions specified by the action
\begin{equation}
    S_\lambda(x) = \frac 1 2 x^2 + \frac{\lambda}{4} x^4\text.
\end{equation}
The relevant learning problem is to estimate $\lambda^{*}$ given samples from $S_{\lambda^{*}}$, for some fixed $\lambda^{*}$. The variance of any estimator we provide will depend on the precise value of $\lambda^{*}$. To make calculations explicit, we select $\lambda^{*} = 0$. The first two non-trivial Schwinger-Dyson equations read
\begin{align}
1 &= \langle x^2\rangle + \lambda \langle x^4\rangle\label{eq:app-sd-1}\text{, and}\\
3 \langle x^2 \rangle &= \langle x^4\rangle + \lambda \langle x^6\rangle\label{eq:app-sd-2}\text.
\end{align}
The second of these relations is the one that yields the score-matching estimator:
\begin{equation}
    \widehat{\lambda}_{\mathrm{SM}} =
    \frac
    {3 \langle x^2 \rangle - \langle x^4 \rangle}
    {\langle x^6\rangle}\text.
\end{equation}
However, by taking linear combinations, these two Schwinger-Dyson relations together define a one-parameter family of estimators of $\lambda^{*}$. Parameterizing by a real number $u$, these estimators are
\begin{equation}
    \widehat{\lambda}_u
    =
    \frac
    {u + (3-u) \langle x^2 \rangle - \langle x^4\rangle}
    {u \langle x^4 \rangle + \langle x^6\rangle}\text.
\end{equation}
This parameterization has been chosen so that $u=0$ yields the score-matching estimator: $\widehat{\lambda}_0 = \widehat{\lambda}_{\mathrm SM}$.

We will now compute the standard deviation $\sigma(u)$ of the estimator $\lambda_u$. Because $\lambda^{*} = 0$, all expectation values involved may be computed immediately by Wick's theorem. The covariance of the random variables $x^2$ and $x^4$ is given (in the basis $\{x^2,x^4\}$) by
\begin{equation}
    \Sigma
    = \left(\begin{matrix}
        2 & 12\\12 & 96
    \end{matrix}\right)
    \text.
\end{equation}
The corresponding gradient of the estimator function $\widehat{\lambda}_u(\langle x^2\rangle,\langle x^4\rangle)$ is
\begin{equation}
    \nabla \widehat\lambda_u = \left(\begin{matrix}
        \frac{\partial\widehat\lambda_u}{\partial\langle x^2\rangle}\\
        \frac{\partial\widehat\lambda_u}{\partial\langle x^2\rangle}
    \end{matrix}
    \right)
    = \frac{1}{3u+15}\left(\begin{matrix}
        3-u\\
        -1
    \end{matrix}
    \right)\text.
\end{equation}
Finally we obtain the variance of the estimator $\widehat\lambda_u$, for the case $\lambda^{*} = 0$, in the limit of a large number of samples $N$:
\begin{equation}
    N^{1/2} \sigma(u)^2
    \approx (\nabla \widehat\lambda_u)^T \Sigma (\nabla \widehat\lambda_u)
    =
    \frac{2 u^2 + 12u + 42}{(3u+15)^2}
    \text.
\end{equation}
It is apparent that $u=0$ is not the minimum, and therefore that the information in Eq.~(\ref{eq:app-sd-1}) is useful in reducing the variance of the learning process below what is obtained by score matching alone.

Of course there is no reason to limit oneself to using a \emph{single} Schwinger-Dyson relation. As shown in the body of this paper an arbitrary number of Schwinger-Dyson relations can be used to construct a loss function, and the resulting estimator will have still-lower variance than may be obtained simply by optimizing $u$.

\section{Grassmann variables}\label{app:grassmann}

Let $\eta_i$ for $i=1,\ldots,n$ be Grassmann variables. Any two Grassmann variables anticommute: $\eta_i \eta_j = -\eta_j \eta_i$. It follows that Grassmanns are nilpotent: $\eta_i^2 = 0$. There are $2^n$ distinct monomials of $n$ Grassmann variables.

A function of Grassmann variables is defined as an arbitrary $\mathbb C$-linear combination of Grassmann monomials. Thus the space of functions on $n$ Grassmann variables is a vector space of dimension $2^n$.
For a function $f(\eta)$, we may define the exponential via Taylor expansion:
\begin{equation}
    e^{f(\eta)}
    = \sum_k \frac{1}{k!} f(\eta)^k
    \text.
\end{equation}
There are a finite number of terms in this expression: For $k > 2^n$, all $f(\eta)^k$ necessarily vanish.

Differentiation and integration act in precisely the same way on Grassmann variables, and are defined by:
\begin{equation}
    \frac{d}{d\eta_i} \eta_j = \int d\eta_i\,\eta_j = \delta_{ij}
    \text.
\end{equation}
Their action on more complicated expressions are defined by linearity. Note that the infinitesimals also anticommute, so that (for example) $\int \!d\eta_1 d\eta_2\, \eta_1 \eta_2 = -1$.

\section{Schwinger-Dyson Relations for Massive Thirring Model}
\label{sec:SD_Th}
Consider the action for the massive Thirring model:
\begin{equation}
    S_{\mathrm{Th}} =
	\sum_{x,x'}\bar{\chi}_x (M_{xx'} + m \delta_{x x'})\chi_{x'}
    +
    \frac{g^2}{2}\sum_{x,\nu}j_{\nu}^2(x)    \text,
\end{equation}
where the massless staggered Dirac operator is defined in Eq.~\eqref{eq:masslessDirac} and the current $j_\nu(x)$ is defined in Eq.~\eqref{eq:current}. The square of the current expands to give a four-Fermi interaction: $
    j_{\nu}^2(x) = -\frac{1}{2}\bar{\chi}_{x+\nu}\chi_{x+\nu}\bar{\chi}_{x}\chi_{x}$.
We will define Schwinger-Dyson relations for operators $Q^{(1)}_{ab} = \bar\chi_{a}\chi_{b}$ and $Q^{(2)}_{a\rho} = \bar{\chi}_{a+\rho}\chi_{a+\rho}\bar{\chi}_{a}\chi_{a}$. For $Q \in \{Q^{(1)},Q^{(2)}\}$, the relations are
\begin{align}
    \langle\partial_{\chi_k}\partial_{\bar{\chi}_k}Q\rangle &= -\langle (\partial_{\bar{\chi}_k}Q)\cdot(\partial_{{\chi}_k}S)\rangle 
\end{align}
\begin{widetext}
Define $X_{a} \equiv \bar{\chi}_a \chi_a, X_{ab} \equiv \bar{\chi}_a \chi_b$, $K_{ab,c} = X_{ab}X_c$, $K_{a,bc} = X_{a}X_{bc}$ and $\Gamma_{abc} \equiv  X_a X_b X_c$. The Schwinger-Dyson relations read:
\begin{align}
    \sum_{i}\bigl(m\delta_{ia}+M_{ia}\bigr) \langle X_{ib}\rangle + \lambda\sum_{\nu}(\langle K_{ab,{a+\nu}} + K_{ab,{a-\nu}}\rangle)&= \delta_{ab}\;, \label{eq:sd1}\\[6pt]
    \sum_i(m\delta_{ia}+M_{ia})\langle K_{ia,a+\rho}\rangle+(m\delta_{i,a+\rho}+M_{i,a+\rho})\langle K_{a,i\, a+\rho} \rangle
    + \lambda\sum_{\nu}(\langle\Gamma_{a,{a+\rho},{a\pm\nu}}\rangle
  + \langle\Gamma_{a,{a+\rho},{a+\rho\pm\nu}}\rangle) &= \langle X_{a+\rho}\rangle + \langle X_{a}\rangle\;.\label{eq:sd2}
\end{align}
For a fixed auxiliary field configuration $A$, the fermionic expectation values follow from Wick's theorem:
\begin{align}
    \langle X_{ab}\rangle \equiv \langle\bar\chi_a\chi_b\rangle = D[A]^{-1}_{ba}\qquad
    \langle X_a X_b\rangle = D[A]^{-1}_{aa}D[A]^{-1}_{bb}-
 D[A]^{-1}_{ab}D[A]^{-1}_{ba}\text,
\end{align}
where the propagator $D[A,g]$ is defined in Eq.~\eqref{eq:staggeredDirac_g}. The six-point functions $\langle\Gamma_{abc}\rangle$
are computed analogously. Full expectation values are then obtained by averaging over the ensemble of configurations. 
Substituting these into Eqs.~\eqref{eq:sd1} and~\eqref{eq:sd2} yields a linear system in the parameters $m$ and $\lambda$, which can be solved to recover the bare couplings.

\end{widetext}

\bibliography{refs}

\end{document}